\documentclass[onecollarge,natbib]{svjour2}
\bibpunct{[}{]}{;}{n}{}{,} 
\smartqed  
\usepackage{graphicx}
\newcommand{\BE}{\begin{equation}}
\newcommand{\EE}{\end{equation}}
\newcommand{\BA}{\begin{eqnarray}}
\newcommand{\EA}{\end{eqnarray}}
\newcommand{\bit}[1]{\mbox{\boldmath$#1$}}
\newcommand{\nn}{\nonumber}
\def\oneht{\textstyle{1\over 2} }

\journalname{Few-Body Systems}

\begin{document}

\title{Goldstone's Theorem on a Light-Like Plane}

\author{Silas R.~Beane}

\institute{Department of Physics, University of Washington, Box 351560\\
              Tel.: 1-206-543-3256,  \email{silas@uw.edu}           
}

\vspace{-0.11in}
\date{Received: date / Accepted: date}

\maketitle

\vspace{-0.11in}
\begin{abstract}
  I review various aspects of chiral symmetry and its spontaneous
  breaking on null planes, including the interesting manner in which
  Goldstone's theorem is realized and the constraints that chiral
  symmetry imposes on the null-plane Hamiltonians. Specializing to QCD
  with $N$ massless flavors, I show that there is an interesting limit
  in which the chiral constraints on the null-plane Hamiltonians can
  be solved to give the spin-flavor algebra $SU(2N)$, recovering a
  result originally found by Weinberg using different methods.
\keywords{Null-plane dynamics \and Chiral symmetry \and Goldstone's theorem}
\end{abstract}

\vspace{-0.11in}
\section{Introduction}
\label{intro}

\noindent 
Understanding of QCD as the correct underlying theory of the strong
interaction relies on the interpretation provided by the parton model,
which can be formulated in a frame-independent manner by quantizing
QCD on a null plane or light
front~\cite{Brodsky:1997de,Belitsky:2005qn}. However, until recently,
a missing link in the null-plane description has been the lack of a
model-independent description of spontaneous chiral symmetry breaking,
which one would think is an essential non-perturbative ingredient in
the matrix elements of local (and nonlocal) operators that appear in
the partonic description.  In particular, a formulation of Goldstone's
theorem that does not depend on the formation of symmetry-breaking
condensates has been lacking. And the manner in which fundamental QCD
relations like the Gell-Mann-Oakes-Renner relation, which involve
symmetry breaking condensates, are realized on the null-plane has
caused considerable confusion. Recent work~\cite{Beane:2013oia} has
clarified these issues and several aspects of chiral symmetry breaking
on null-planes will be reviewed here.

Chiral constraints on low-energy QCD observables are usually viewed in
the context of chiral perturbation theory, which through fluctuations
of the chiral condensate, allows one to calculate the long-distance
Goldstone boson contributions to hadronic observables in a
perturbative expansion in Goldstone boson masses and momenta. However,
as pointed out by Weinberg long ago~\cite{Weinberg:1969hw}, in
addition to these ``dynamical'' consequences of chiral symmetry in the
infrared that are accessed using chiral perturbation theory, there are
additional consequences of chiral symmetry that are ``algebraic'' in
nature, and are generally expressed as sum rule constraints on
observables that arise from specific assumptions about the asymptotic
behavior of scattering amplitudes. The null-plane chiral constraints
that are discussed here are precisely of the ``algebraic'' type found
by Weinberg who used a completely different viewpoint that does not rely
on null-plane quantization. This is gratifying since, at the end of
the day, physics does not depend on the particular frame, coordinates
or quantization surfaces that are chosen.

\vspace{-0.11in}
\section{Null-Plane Hamiltonians}
\label{sec:1}

\noindent
In the usual instant-form quantization, the four dynamical generators
are the energy, or time-evolution operator, and the boosts. However,
as the boosts are not associated with any observable, in the
instant-form, the focus is on obtaining the Hamiltonian energy
operator. By contrast, on a null-plane there are three dynamical
generators, corresponding to the light-cone time evolution operator,
or energy, and the transverse components of the spin
operator~\cite{Dirac:1949cp}.  As the spin is an important observable,
we refer to the three dynamical generators as
Hamiltonians. Remarkably, on a light-like plane, all dynamical
information of a Poincar\'e invariant theory of quantum mechanics is
contained in the three reduced Hamiltonians $M\,\!{\cal J}_r$ (with
$r=1,2$) and $M^2$ which encode the spin content and spectrum,
respectively~\cite{Leutwyler:1977vy}. These reduced Hamiltonians
commute with six of the kinematical generators of the Poincar\'e
algebra and satisfy the $U(2)$ algebra together with the seventh
kinematical generator, ${\cal J}_3$:
\BA
[\, {\cal J}_3 \, ,\,  M\;\!\!{\cal J}_r  \, ] \ =\   i\,\epsilon_{rs} M\;\!\!{\cal J}_s
\ \  \ & , &\  \ \ 
[\, {\cal J}_3 \, ,\,  M^2  \, ]\, = \, 0 \ ; \nonumber \\
{} [\, M\;\!\!{\cal J}_r \, ,\,  M\;\!\!{\cal J}_s  \, ] \ =\  i\,\epsilon_{rs} M^2 {\cal J}_3  
\ \ \ & ,& \ \ \
[\, M^2 \, ,\,  M\;\!\!{\cal J}_r  \, ]\, = \, 0 \ .
\label{eq:dynalgarbB}
\EA 
This decoupling of the kinematical and dynamical generators is an
important property of null-plane quantization, particularly as regards
the issue of chiral symmetry breaking, as we know that all chiral
symmetry breaking is necessarily contained in the three reduced
Hamiltonians. 

As spin is dynamical on the null-plane, {\it a priori} one does not
expect that the null-plane description in terms of Lagrangian field
theory will exhibit manifest Lorentz invariance. Spin takes its proper
form only when interactions are explicitly taken into account --that
is, when the dynamics of the system are fully solved.  Hence the
arrangement of the spectrum of the theory into representations of the
Lorentz group is evident only in the solution of the theory.  In
particular, while the spin of the system is usually given by the sum
of the spins of the constituents, here that is no longer the case. The
spin of the system is given by the sum of the spins of the
constituents as well as by the interactions among the constituent
spins.

\vspace{-0.11in}
\section{Null-Plane Chiral Symmetry}
\label{sec:2}

\noindent
Consider a system whose action has an exact $SU(N)_R\otimes SU(N)_L$ symmetry,
with Noether currents given by ${\tilde J}_\alpha^\mu(x)$ and ${\tilde
  J}_{5\alpha}^\mu(x)$, and whose internal charges, defined by
\begin{eqnarray}
{\tilde Q}_\alpha \ &=& \ \int\, d x^-\, d^2 \bit{x}_\perp\, {\tilde J}^+_{\alpha}(x^-, {\vec x}_\perp) \ ; \\
{\tilde Q}^5_\alpha \ &=& \ \int\, d x^-\, d^2 \bit{x}_\perp\, {\tilde J}^+_{5\alpha}(x^-, {\vec x}_\perp) \ ,
\label{eq:npchargesgenB}
\end{eqnarray}
satisfy the chiral algebra:
\BA
[\, {\tilde Q}^\alpha\,  ,\, {\tilde Q}^\beta\, ]\, =\, i\,f^{\alpha\beta\gamma} \, {\tilde Q}^\gamma
\ & , &\  
[\, {\tilde Q}_5^\alpha\, ,\, {\tilde Q}^\beta\, ]\, =\,  i\,f^{\alpha\beta\gamma}\,  {\tilde Q}_5^\gamma \ ;
\nn \\
&&\!\!\!\!\!\!\!\!\!\!\!\!\!\!\!\!\!\!\!\!\!\!\!\!\!\!\!\!\!\!\!\!\![\, {\tilde Q}_5^\alpha \, ,\, {\tilde Q}_5^\beta\, ]\,  =\,  
i\,f^{\alpha\beta\gamma} \, {\tilde Q}^\gamma \ .
\label{eq:LCalgb}
\EA
Here we assert that on a null-plane both types of chiral charges annihilate the vacuum. That is,
\BA
{\tilde Q}^\alpha\, |\, 0\,\rangle\, =\, {\tilde Q}_5^\alpha\, |\, 0\,\rangle\, =\, 0 \ .
\label{eq:vacuumgen}
\EA It is straightforward to verify these relations explicitly in
null-plane QCD~\cite{Beane:2013oia}. In general, this structureless
nature of the null-plane vacuum arises because the null-plane momentum
operator has a spectrum confined to the open positive half-line.
(Here one should keep in mind that the correct treatment of the
singularities that arise in null-plane quantization is a subtle mathematical
issue~\cite{Jegerlehner:1974re}.) 
The important consequence of this is that the null-plane vacuum is
invariant with respect to the full $SU(N)_R\otimes SU(N)_L$ symmetry,
even in the phase in which the symmetry is said to be spontaneously
broken. In particular, this implies that there can be no vacuum
condensates that break $SU(N)_R\otimes SU(N)_L$ on a
null-plane~\cite{Beane:2013oia}.  If the chiral symmetry is not
spontaneously broken then one expects that the chiral currents are
conserved and the chiral charges commute with the reduced
Hamiltonians.  However, if the symmetry is spontaneously broken, then
one must have
\BA
[\, {\tilde Q}^\alpha_5\, ,\, M^2\, ] \ \neq \ 0 \quad ; \quad
[\, {\tilde Q}^\alpha_5\, ,\, M {\cal J}_\pm\, ]\ \neq \ 0 \ , 
\label{eq:Qcomm2agen}
\EA
where ${\cal J}_\pm\equiv{\cal J}_1\pm i{\cal J}_2$.
That is, given the invariance of the vacuum, symmetry breaking must
be present in the Hamiltonians. The particular pattern of breaking depends on the assumed form of chiral symmetry breaking.
In QCD with $N$ flavors of massless quarks the Lie brackets among the reduced Hamiltonians and the chiral charges 
are easily evaluated to give~\cite{Beane:2013oia}
\BA
{\cal P}^{\alpha\beta ;\mu\nu}\;[{\tilde Q}^\mu_5\, ,\, [\, {\tilde Q}^\nu_5\, ,\, M^2]]\ =\ {\cal P}^{\alpha\beta ;\mu\nu}\;[{\tilde Q}^\mu_5\, ,\, [\, {\tilde Q}^\nu_5\, ,\, M {\cal J}_\pm]]\ =\ 0 \ ,
\label{eq:fundmix}
\EA 
where
\BA {\cal P}^{\alpha\beta ; \mu\nu}\ \equiv\
\delta^{\alpha\nu}\delta^{\beta\mu}\ -\
\frac{1}{N^2-1}\;\delta^{\alpha\beta}\delta^{\mu\nu} \ -\
\frac{N}{N^2-4}\;d^{\alpha\beta\gamma}d^{\mu\nu\gamma} \ .  
\EA 
These constraints imply that $M^2$ and $M {\cal J}_\pm$ transform as linear
combinations of $(1,1)$, $(\bar{\bf N},{\bf N})$ and $({\bf
  N},\bar{\bf N})$ representation of $SU(N)_R\otimes SU(N)_L$.

\vspace{-0.11in}
\section{Goldstone's Theorem}
\label{sec:3}

\noindent 
The standard field-theoretic paradigm tells us that a classical symmetry of an action
has three possible fates after quantization: the symmetry remains
unbroken and the currents are conserved, the symmetry is spontaneously
broken and once again the currents are conserved, or the symmetry is
anomalous and the associated current is not conserved. The null plane
realizes a fourth possibility: the symmetry is spontaneously broken
and the associated current is not conserved.  This pattern is a necessary
consequence of the vacuum being invariant with respect to all internal
symmetries on a null plane.  Now if we add an explicit chiral symmetry
breaking operator to the action, then, in general, one has
\BA
\partial_\mu {\tilde J}^\mu_{5\alpha}(x^-, {\vec x}_\perp, x^+) \ = \ \epsilon_{\chi}\ {\tilde P}_{\alpha}(x^-, {\vec x}_\perp, x^+) \ ,
\label{eq:genpcac}
\EA
where $\epsilon_{\chi}$ is a parameter that measures the amount of explicit chiral symmetry breaking that is present in the Lagrangian.
Using the short hand,
\BA
|\,h\,\rangle\equiv \,|\, p^+\, ,\, {\vec p}_\perp\,;\, \lambda\,,\, h\,\rangle \ ,
\label{eq:HME2}
\EA for the momentum eigenstates, we take the matrix element of eq.~\ref{eq:Qcomm2agen} (left side)
between momentum eigenstates, which gives~\cite{Beane:2013oia}
\BA
\langle\, h'\, |\lbrack\, {\tilde Q}^5_\alpha(x^+) \, ,\, M^2\rbrack |\, h\, \rangle\, &=&\, -2i\,p^+\,
\epsilon_{\chi}\, \int\, d x^-\, d^2 \bit{x}_\perp\, \langle\, h'\, | {\tilde P}_{\alpha}(x^-, {\vec x}_\perp, x^+) |\, h\, \rangle \ . 
\label{eq:mixedchiral1}
\EA 
If the right hand side of this equation vanishes for all $h$ and $h'$,
then there can be no chiral symmetry breaking of any kind, since ---as we
have argued--- there can be no symmetry breaking condensates and therefore all symmetry
breaking must reside in the Hamiltonians. Hence, in order that the
chiral symmetry be spontaneously broken, the chiral current cannot be
conserved and we have the following
constraint~\cite{Kim:1994rm,Yamawaki:1998cy} in the limit where the
explicit symmetry breaking is turned off, $\epsilon_\chi\rightarrow
0$:
\BA \int\, d x^-\, d^2 \bit{x}_\perp\, \langle\, h'\, |{\tilde
  P}_{\alpha}(x^-, {\vec x}_\perp, x^+)|\, h\, \rangle
&\longrightarrow & \frac{1}{\epsilon_\chi}\ +\ \ldots \
,\label{eq:pizeromodea} 
\EA 
where the dots represent other terms that are regular in the limit
$\epsilon_\chi\rightarrow 0$.  Now we will show that this condition
implies the existence of $N^2-1$ Goldstone bosons. As in the
instant-form formulation of chiral symmetry breaking, we can treat the
operator ${\tilde P}_\alpha$ as an interpolating operator for
Lorentz-scalar fields $\phi_\alpha^i$ that carry the same quantum
numbers, and we can write
\BA
{\tilde P}_\alpha(x) \ =\ \sum_i {\cal Z}_i\,\phi_\alpha^i(x) \ 
\label{eq:zfactors}
\EA
where the ${\cal Z}_i$ are overlap factors. 
Now we can use standard technology, i.e. the reduction formula,
to relate the matrix elements of this operator between
physical states to transition amplitudes. 
The S-matrix element for the transition $h(p)\rightarrow h'(p')+\phi^i_\alpha(q)$ is defined as
\BA
\langle \, h'\,;\, \phi_\alpha^i(q)\, |
S |\, h\, \rangle  &\equiv&  i(2\pi)^4\,\delta^4(\, p\,-\, p'-q)\,
{\cal M}^i_\alpha (\,p',\,\lambda',\, h'\,;\,p,\,\lambda,\, h\,) \ ;\nn \\
&=& i\int d^4x\,e^{-iq\cdot x}\,\left(-q^2+M_{\phi^i}^2\right)\,\langle\, h'\, |\, \phi_\alpha^i(x)\,|\, h\, \rangle
\label{eq:HME3}
\EA
where ${\cal M}^i_\alpha$ is the Feynman amplitude and in the second line the reduction
formula has been used. It follows that
\BA
\langle\, h'\, |\, \phi_\alpha^i(x)\,|\, h\, \rangle & =& 
-e^{iq\cdot x}\,{1\over{q^2-M_{\phi^i}^2}}\, {\cal M}^i_\alpha(q) \ .
\label{eq:reductiongen}
\EA
Using this result, together with eq.~\ref{eq:zfactors}, in eq.~\ref{eq:mixedchiral1} gives
\BA
\langle\, h'\, |\lbrack\, {\tilde Q}^5_\alpha(x^+) \, ,\, M^2\rbrack |\, h\, \rangle &=& 2i\,p^+\,
(2\pi)^3\,\delta(\,q^+\,)\,\delta^2(\,{\vec q}_\perp \,)\,e^{ix^+q^-}\sum_i \frac{\epsilon_\chi\, {\cal Z}_i}
{2q^+q^-\,-\,{\vec q}_\perp^{\;2}-M_{\phi^i}^2}\,{\cal M}^i_\alpha(q) \ ; \nn \\ 
&=&{} -2i\,p^+\, (2\pi)^3\,\delta(\,q^+\,)\,\delta^2(\,{\vec q}_\perp \,)\,e^{ix^+q^-}\sum_i \frac{\epsilon_\chi\, {\cal Z}_i}
{M_{\phi^i}^2}\,{\cal M}^i_\alpha(q^-) \ .
\label{eq:mixedchiralproof}
\EA 
In order that the right hand side of this equation be non-zero in the symmetry limit,
there must be at least one field $\phi_\alpha^i$ whose mass-squared
vanishes proportionally to the symmetry-breaking parameter
$\epsilon_\chi$ as $\epsilon_\chi\rightarrow 0$. We will denote this
field as $\pi^\alpha\equiv \phi_\alpha^1$ with
\BA M_\pi^2 \ =\ c_p\,
\epsilon_\chi \ ,
\label{eq:pionmass}
\EA 
where $c_p$ is a constant. There are therefore
$N^2-1$ massless fields $\pi_\alpha$ in the symmetry limit, which are
identified as the Goldstone bosons.  It is important to note that this proof
of Goldstone's theorem relies entirely on physical matrix elements.
That is, unlike the usual textbook proof of Goldstone's theorem,  there is no need here
to assume the existence of vacuum condensates that transform non-trivially with respect
to the chiral symmetry group. Of course, in instant-form quantized QCD, we know that the
proportionality constant $c_p$ in eq.~\ref{eq:pionmass} contains the quark
condensate. We will return to this point in the next section.

Writing ${\tilde P}_\alpha \ =\ {\cal Z}\,\pi_\alpha\ +\ \ldots$
where the dots represent other (non-Goldstone) boson fields, and 
\BA
\langle\, h'\, |\,  \partial_\mu\,{\tilde J}^\mu_{5\alpha}(x)\,|\, h\, \rangle & =& 
\langle\, h'\, |\,  {\bar{\cal Z}}\,M_\pi^2\,\pi_\alpha(x)\,|\, h\, \rangle \ ,
\label{eq:pcacgen}
\EA
where ${\bar{\cal Z}}\equiv{\cal Z}/c_p$. It is now a standard exercise to determine the overlap factor. Defining
the Goldstone-boson decay constant, $F_\pi$, via
\BA
\langle\, 0\, |\,{\tilde J}^\mu_{5\alpha}(x)\,|\, \pi_\beta\, \rangle & \equiv& -i\,p^\mu\;F_\pi\,\delta_{\alpha\beta}\;e^{ip\cdot x} \ ,
\label{eq:piondecaydef}
\EA
where $|\, \pi_\beta\, \rangle \equiv |\, p^+\, ,\, {\vec p}_\perp\,;\, 0,\, \pi_\beta\, \rangle$,
one finds  $\bar{\cal Z}=F_\pi$.

\vspace{-0.11in}
\section{Null-Plane Condensates}
\label{sec:4}

\noindent In QCD with $N$ degenerate flavors of quarks eq.~\ref{eq:pionmass}
takes the form of the Gell-Mann-Oakes-Renner formula~\cite{GellMann:1968rz}
\BA
-{{M}\,\langle\, \Omega \, |\,\bar\psi\psi  \,|\, \Omega\, \rangle}\ =\ \oneht\,N\,M_\pi^2\,F_\pi^2 \ \ +\ \dots \ ,
\label{eq:ifgmor}
\EA
where $M$ is the quark mass, $\psi$ is the quark field, and $|\, \Omega\, \rangle$ is the instant-form vacuum.
With respect to the instant-form chiral charges, ${Q}^\alpha_5$, $\bar\psi\psi$ transforms as
$(\bar{\bf N},{\bf N})\oplus({\bf N},\bar{\bf N})$ under $SU(N)_R\otimes SU(N)_L$. 
It is a simple textbook exercise to check this; one finds 
\BA
[\, {Q}^\alpha_5\, ,\, \psi \, ]\, =\,  -\gamma_5\,T^\alpha\,\psi \ ,
\EA
where the $T_\alpha$  are $SU(N)$ generators. This transformation property implies that the
quarks transform {\it irreducibly} with respect to $SU(N)_R\otimes SU(N)_L$; that is,
$\psi_{R} \in  ({\bf 1},{\bf N})$ and $\psi_{L}^\dagger  \in  (\bar{\bf N},{\bf 1})$.
The claimed transformation property of $\bar\psi\psi$ then follows.

How then is this relation reconciled with our claim that there are no symmetry-breaking condensates on a null-plane?
With respect to the null-plane (tilded) charges, one finds~\cite{Carlitz:1974sg}
\BA [\, {\tilde Q}^\alpha_5\, ,\, \psi \, ]\, =\,
-\gamma_5\,T^\alpha\,\psi \ -\
i\;\gamma_5\;\gamma^+\;T^\alpha\;\frac{1}{\partial^+_{M}} \psi \ ,  
\label{eq:psitrans}
\EA 
from which it follows that 
\BA
\psi_{R}\, ,\, \psi_{L}\ \in \ ({\bf 1},{\bf N})\oplus ({\bf N},{\bf 1}) \quad , \quad
\psi_{R}^\dagger\, ,\,\psi_{L}^\dagger \ \in \ ({\bf 1},\bar{\bf
  N})\oplus(\bar{\bf N},{\bf 1}) \ .  \EA 
That is, the quarks transform {\it reducibly} with respect to $SU(N)_R\otimes SU(N)_L$.
Because of this reducibility, the bilinear $\bar\psi\psi$ always contains the 
$SU(N)_R\otimes SU(N)_L$ singlet! Indeed, in terms of the dynamical quark field
$\psi_+$ the null-plane expression of the Gell-Mann-Oakes-Renner relation
can be formally written as~\cite{Wu:2003vn}
\BA {{M}\,\langle\, 0\, |\,i\;\bar\psi_+ \gamma^+
  \frac{1}{\partial^+_{M}} \psi_+\,|\, 0\, \rangle}\ =\ \oneht\,
N\,M_\pi^2\,F_\pi^2 \ \ +\ \dots \ ,  
\label{eq:npgoar}
\EA 
where now $|\, 0\, \rangle$ is the null-plane vacuum state,
and the nonlocal operator $1/{\partial^+_{M}}$ is defined in Ref.~\cite{Beane:2013oia}.
Hence we see that a chiral-symmetry breaking condensate in the instant-form
formulation of QCD is replaced by a chiral-symmetry conserving
condensate in the null-plane formulation. Of course both relations,
eq.~\ref{eq:ifgmor} and eq.~\ref{eq:npgoar}, contain precisely the
same physics, as they must.

\vspace{-0.11in}
\section{Weinberg's Recovery of Spin-Flavor Symmetries}
\label{sec:5}

\noindent The goal of finding solutions of the algebraic system that
mixes the chiral charges and the reduced Hamiltonians may seem hampered
by the existence of no-go theorems that forbid the non-trivial
mixing of space-time and internal symmetries. In the null-plane
formulation, these no-go theorems are avoided because it is only the
dynamical part, i.e. the Hamiltonians, of the null-plane Poincar\'e algebra that
mix with the chiral symmetry generators~\cite{Leutwyler:1977vz}.

While a general solution of the null-plane QCD operator algebra,
given by eqs.~\ref{eq:dynalgarbB}, \ref{eq:LCalgb}, and \ref{eq:fundmix},
is not known, there is a very-interesting limiting
case in which the algebra yields an important and familiar solution.
Here we will work with the QCD operator algebra.  However, it is
important to keep in mind that matrix elements of the operator
relations between physical, hadronic states must be taken in order to
extract observables.  We first define
\BA
[\, {\tilde Q}^\alpha_5\, ,\, M\, ] \, \equiv \,  \epsilon^\alpha \ ,
\label{epsa}
\EA
and throw away terms of ${\cal O}(\epsilon)$. This implies that all
chiral symmetry breaking must occur in the spin Hamiltonians, $M {\cal J}_\pm$.
In this limit, the QCD operator algebra reduces to
\BA
[\, {\cal J}_i \, ,\,   {\cal J}_j  \, ] \, =\, i\,\epsilon_{ijk}\,  {\cal J}_k
\EA
which generates $SU(2)$ spin, the $SU(N)_R\otimes SU(N)_L$ algebra of 
eq.~\ref{eq:LCalgb}, and the remaining non-trivial Lie bracket mixes the chiral generators and the spin Hamiltonians:
\BA 
{\cal P}^{\alpha\beta ;\mu\nu}\;[{\tilde Q}^\mu_5\, ,\, [\, {\tilde
  Q}^\nu_5\, ,\, {\cal J}_\pm]]\ =\ 0 \ .  
\label{eq:OPQcomm2b}
\EA 
Remarkably, this simplified algebra can be put into a familiar form. Consider
an operator $G_{\alpha i}$, which transforms as an adjoint of $SU(N)$ and as a vector
with respect to rotations. In general, the commutator of $G^{\alpha i}$ with itself may be expressed as
\BA
[\, G_{\alpha i} \, ,\, G_{\beta j} \, ] \, =\, 
i\,f_{\alpha\beta\gamma}\, {\cal A}_{ij,\gamma}\, +\, i\,\epsilon_{ijk}\,  {\cal B}_{\alpha\beta,k} \ ,
\label{eq:su4commgen}
\EA
where ${\cal A}_{ij,\gamma}={\cal A}_{ji,\gamma}$ and ${\cal
  B}_{\alpha\beta,k}={\cal B}_{\beta\alpha,k}$.  Now we identify
$G^{\alpha 3}\ \equiv \ {\tilde Q}^\alpha_5$. Detailed consideration of the 
properties of $G^{\alpha\beta}$ determines  ${\cal A}$ and ${\cal B}$ and finally leads to~\cite{Weinberg:1994tu,Beane:2013oia}
\BA
[\, G_{\alpha i} \, ,\, G_{\beta j} \, ] \, =\, 
i\,\delta_{ij}\,f_{\alpha\beta\gamma}\, {\tilde Q}_\gamma\, +\, \frac{2}{N}\,i\,\delta_{\alpha\beta}\,\epsilon_{ijk}\,{\cal J}_k  \, + \, i\epsilon_{ijk}\,d_{\alpha\beta\gamma}\, {G}_{\gamma k} \ ,
\label{eq:su2N}
\EA
which together with 
\BA &&[\, {\tilde Q}_\alpha\, ,\, G_{\beta i}\, ]\, =\,
i\,f_{\alpha\beta\gamma} \, G_{\gamma i} \ \ \ , \ \ \
[\, {\cal J}_i \, ,\,   G_{\alpha j}  \, ] \, =\, i\,\epsilon_{ijk}\,  G_{\alpha k} \ ; \\
&&[\, {\tilde Q}_\alpha\, ,\, {\tilde Q}_\beta\, ]\, =\,
i\,f_{\alpha\beta\gamma} \, {\tilde Q}_\gamma \ \ \ , \ \ \ [\, {\cal
  J}_i \, ,\, {\cal J}_j \, ] \, =\, i\,\epsilon_{ijk}\, {\cal J}_k
\EA 
close to the algebra of the group $SU(2N)$. 
It is important to stress that this $SU(2N)$ symmetry 
is truly a dynamical symmetry; it is unrelated to the 
invariance of QCD in the non-interacting limit.

Importantly, this symmetry can be viewed as emerging in a particular
limit of QCD.  As the hadronic matrix element of eq.~\ref{epsa},
$\langle\,h'\,|\epsilon^\alpha|\, h\,\rangle\sim M_h-M_{h'}$, and
baryons within large-$N_c$ multiplets have mass splittings that scale
as $1/N_c$~\cite{Witten:1979kh}, standard large-$N_c$ QCD scaling
rules suggest that for baryons $\epsilon^\alpha\sim 1/N_c$.  Moreover,
as the matrix element of chiral charges between baryon states scales
as $N_c$, the $SU(2N)$ symmetry formally reduces to the contracted
$SU(2N)$ symmetry~\cite{Weinberg:1994tu} for baryons in the large-$N_c$ limit,
as must occur on general
grounds~\cite{Gervais:1983wq,Dashen:1993as,Dashen:1993jt}.

It is instructive to consider a simple example in order to see
how null-plane chiral symmetry constrains hadronic structure. Consider the case
$N=3$. Using the null-plane chiral transformation properties of the quarks~\cite{Beane:2013oia}
\BA
\psi_{+R}\ =\ \psi_{+\uparrow}\ &\in& \ ({\bf 1},{\bf N}) \qquad , \qquad \psi_{+R}^\dagger \ =\ \psi_{+\downarrow}^\dagger\ \in \ ({\bf 1},\bar{\bf N})\ ; \\
\psi_{+L}\ =\ \psi_{+\downarrow}\ &\in& \ ({\bf N},{\bf 1}) \qquad , \qquad \psi_{+L}^\dagger \ =\ \psi_{+\uparrow}^\dagger\ \ \in \ (\bar{\bf N},{\bf 1})\ ,
\label{eq:chiralassign}
\EA one sees that a $\lambda=3/2$ baryon-like operator
$\psi_{+\uparrow}\psi_{+\uparrow}\psi_{+\uparrow}$ transforms as
$({\bf 1},{\bf 1})$, $({\bf 1},{\bf 8})$, or $({\bf 1},{\bf 10})$ with
respect to $SU(3)_R\otimes SU(3)_L$. Therefore, if the baryon is a
decuplet of $SU(3)_F$ with its $\lambda=3/2$ part in the $({\bf
  1},{\bf 10})$, then one readily checks that its $\lambda=1/2$
component must transform as $({\bf 3},{\bf 6})$ or $({\bf 6},{\bf
  3})$.  However, the various helicity states are unrelated by chiral
symmetry in itself. Strangely, it is the mixed Lie-bracket,
eq.~\ref{eq:OPQcomm2b}, the contribution of spontaneously broken
chiral symmetry to the spin Hamiltonian, that relates the helicities!
It is counter intuitive to have states that are unrelated in the
symmetry limit, become part of a symmetry multiplet in the
broken-symmetry limit. Nevertheless, this is precisely what occurs on
the null-plane.  Taking the $\lambda=1/2$ baryon decuplet to transform
as $({\bf 3},{\bf 6})$ together with an baryon octet spin-$1/2$, and
with their negative-helicity partners in $({\bf 10},{\bf
  1})\oplus({\bf 6},{\bf 3})$, together form the ${\bf
  56}$-dimensional representation of $SU(6)$, which is the familiar
ground-state baryon assignment in the non-relativistic quark
model. The difference between what has been found here and the quark
model is that the symmetry that arises from the null-plane algebra
follows directly from QCD symmetries and their pattern of breaking.
In particular, this symmetry has nothing to do with the
non-relativistic limit.

In summary,  beginning from the null-plane QCD operator algebra, the 
assumption that the chiral-symmetry breaking part of the null-plane reduced Hamiltonian, $M^2$,
is small, implies all of the usual consequences of the non-relativistic quark model, but without the need to
assume constituent quark degrees of freedom~\cite{Weinberg:1994tu}.
The next step is to investigate how the ${\cal O}(\epsilon)$ terms that were
neglected modify the simple quark-model-like picture. For instance, it is clear that
the nucleon null-plane wavefunction will then be a mixture of various representations
which will {\it inter alia} constrain the spin content and spectrum of the light
baryons.

\begin{acknowledgements}
This work was supported in part by NSF continuing grant PHY1206498.
\end{acknowledgements}

\end{document}